\documentclass[12pt]{article}
\usepackage{amsmath}
\usepackage{amssymb}

\newtheorem{thm}{Theorem}[section]
\newtheorem{prop}[thm]{Proposition}
\newtheorem{lem}[thm]{Lemma}
\newtheorem{dfn}[thm]{Definition}
\newtheorem{cor}[thm]{Corollary}
\newtheorem{con}[thm]{Conjecture}
\newtheorem{rem}[thm]{Remark}

\def\bb{\begin{eqnarray}}
\def\ee{\end{eqnarray}}
\def\ds{\displaystyle}

\newcommand{\qed}{\hfill \fbox{}\medskip}
\newcommand{\proof}{\medskip\noindent{\bf Proof.}\quad }

\title{Periodic ILW equation with discrete Laplacian}

\author{Jun'ichi Shiraishi\footnote{Graduate School of 
Mathematical Sciences, The University of Tokyo, 3-8-1 Komaba Meguro-ku Tokyo 153-8914, Japan} and 
Yohei Tutiya\footnote{Ohara Graduate School of Accounting, 2-2-10 Nishi-Kanda Chiyoda-ku Tokyo 101-0065, Japan}}

\date{}

\begin{document}

\maketitle

\begin{abstract}
We study an integro-differential equation which generalizes 
the periodic intermediate long wave (ILW) equation. The kernel of the singular integral involved is
an elliptic function written as 
a second order difference of the Weierstrass $\zeta$-function. 
Using Sato's formulation, we show the integrability and construct
some special solutions. An elliptic solution is also obtained.
We present a conjecture based on a Poisson structure that it gives an alternative description of 
this integrable hierarchy. We note that this Poisson algebra in turn is related to 
a quantum algebra related with 
the family of Macdonald difference operators.

\end{abstract}

\section{Introduction}

In this paper, we consider an integrable differential equation 
with a singular integral term associated with a doubly periodic function. 
We often classify the known integrable equations with singular integrals
according to the periodicity of the kernel functions. 
Namely, the Benjamin-Ono equation \cite{B,O} corresponds to the case with no period (rational function) 
because it has the Hilbert transformation, 
and the intermediate long wave (ILW) equation \cite{Jo,Ku} corresponds to 
the singly periodic case (trigonometric function) as
hyperbolic cotangent is involved as the kernel.
The doubly periodic case was introduced in \cite{AFSS}
as a periodic version of ILW equation.

We aim at constructing an integrable equation
which recovers all these equations 
as special limits, and which also relates to the 
theory of the Macdonald polynomials \cite{Mac} (see \cite{S,FHMSWY} also).
We propose
that the kernel of our singular integral is an elliptic function
having simple poles at three points $\gamma,0,-\gamma$ in the fundamental parallelogram 
with residues $1,-2,1$ respectively and holomorphic elsewhere. 

Let $\omega_1,\omega_2$ and $\gamma$ be complex numbers such that
the ratio $\delta=\omega_2/\omega_1$ satisfies ${\rm Im}(\delta)>0$, and 
$0<{\rm Im}(\gamma)<{\rm Im}(\delta)$.
Let $x$ and $t$ be real independent variables, and let $\eta(x,t)$  
be an analytic function satisfying the periodicity condition $\eta(x+1,t)=\eta(x,t)$.
We consider the integro-differential equation
\begin{eqnarray}
\frac\partial{\partial t}\eta(x,t)
=
\eta(x,t)\cdot 
\frac{i \omega_1}{\pi }
\int_{-1/2}^{1/2}
\hspace{-2.2em}\backslash\hspace{.5em}\,\,\,\,\,
(\Delta_\gamma\zeta)(2\omega_1(y-x))\cdot 
\eta(y,t)dy,
\label{ILW}
\end{eqnarray}
where $\zeta(x)=\zeta(x;2\omega_1,2\omega_2)$ denotes the Weierstrass $\zeta$-function 
\cite{WW}  (see (\ref{zeta}) in Appendix A),
the discrete Laplacian $\Delta_\gamma$ is defined by
$(\Delta_\gamma f)(x)=f(x-\gamma)-2f(x)+f(x+\gamma)$,
and the integral $\int\hspace{-0.8em}\backslash\hspace{.5em}$
means the Cauchy principal value.
\bigskip

Our purpose in this paper is to study (\ref{ILW}) from
several viewpoints. First we use the standard method which 
transforms (\ref{ILW}) into a difference-differential form \cite{AFSS,SAK,KAS,STA}.
Then we follow the method developed in \cite{UT,TS}
to utilize the Sato theory to construct an integrable hierarchy which
includes  (\ref{ILW}).
Then we study the system of integrals of motion
associated with (\ref{ILW}) in terms of the Sato theory. 

We remark the following.
In \cite{SAK,STA}, the conserved densities for 
the (periodic) ILW 
was studied by using the B\"acklund transformation.
However, we do not know a B\"acklund transformation for 
(\ref{ILW}) at present. It is an open question to find it and 
compare the approach given in this paper with those classical analysis 
of conserved quantities.

In the papers \cite{Sa,LR}, a Hamiltonian approach to the ILW equation
was pushed forward by using the Gel'fand-Dikij brackets and the bi-Hamiltonian structure. 
We will develop a Hamiltonian description of (\ref{ILW}) in the same spirit as theirs. 
Our situation, however, might be a little tangled in the following sense. 
In one hand, it 
has a direct connection with the Macdonald difference operators \cite{Mac},
or to be more precise, 
to its elliptic analogue defined through the algebra of Feigin and Odesskii \cite{FO} (see \cite{S} 
and \cite{FHMSWY}). On the other hand, we also attempt to connect the Hamiltonian approach 
with the Lax formulation of Sato.
It is a future problem to understand analogues of 
the Gel'fand-Dikij brackets and the bi-Hamiltonian structure for  (\ref{ILW}). 
 
Finally, we make an important comment that the Poisson algebra in this paper has a deep connection with
the one found in \cite{Fr}. 
The difference analogue of $N$-th KdV studied by Frenkel 
has two parameters $q$ and $N$. It can be found that 
if we set  
 $q=e^{2\pi i \gamma}$ and impose the condition $\delta=N\gamma$,
 we almost recovers Frenkel's Poisson algebra, but missing the delta function terms 
 which typically appears in the deformed $\cal W$-algebras (see \cite{FR}).
 \bigskip

This article is organized as follows.
In Section 2, we rewrite (\ref{ILW}) in a form of differential-difference equation.
Then we show the ordinary ILW equation with periodicity can be
obtained from (\ref{ILW}) in the limit $\gamma\rightarrow 0$.
In Section 3, 
by using the standard Sato theory
we present an integrable hierarchy which contains (\ref{ILW}) in the lowest order. 
We study the structure of the integrals of motion in some detail in this setting.
Finally, Section 4 is devoted to an alternative description based on  a Poisson structure 
derived from a quantum mechanical integrable model
associated with the Macdonald theory,
from which 
we recover the same equation  (\ref{ILW}), and presumably all the 
equations given in the hierarchy.

\section{Differential-difference form}
\subsection{integral operator $\mathsf{ T}$}
Let $\mathsf{ T}$ be the integral transformation defined by
\bb
(\mathsf{ T}f)(x)
=\
{i\omega_1\over \pi }
\int_{-1/2}^{1/2}
\hspace{-2.2em}\backslash\hspace{.5em}\,\,\,\,\,
(\Delta_\gamma\zeta)(2\omega_1(y-x))\cdot
f(y)dy,
\label{sekibunhenkan}
\ee
then, (\ref{ILW}) can be written as
$\dot{\eta}=\eta \,(\mathsf{ T}\eta)$.

Decomposing $\eta(x,t)$ by the Plemelj formalism \cite{M},
one obtains a differential-difference equation and a bilinear equation from (\ref{ILW}).
Here and hereafter, we set $p=e^{2\pi i \delta},q=e^{2\pi i \gamma}$ for simplicity,
and  let $D$ be a domain in the complex $z$-plane containing the infinite strip 
$0\leq {\rm Im}(z)\leq {\rm Im}(\delta)$. 

\begin{lem}
For any nonzero integer $m$ we have
\begin{eqnarray}
\mathsf{ T}(e^{2\pi i m x})={(1-q^m)(1-p^mq^{-m})\over 1-p^m}e^{2\pi i m x},
\end{eqnarray}
and $\mathsf{ T}(1)=0$.
\end{lem}
\proof Use (\ref{fou}) in Appendix A. \qed
\begin{cor}\label{cor}
Suppose that an analytic function $f(z)$ on $D$ 
satisfies the periodicity $f(x+1)=f(x)$, then we have
\bb
&&
\mathsf{ T} (f(x)-f(x+\delta))=f(x)-f(x+\gamma)
-f(x+\delta
-\gamma)+f(x+\delta).\label{Tff}
\ee
\end{cor}

Define difference operators $T$ and $S$ acting on the variable $x$
by $Tf(x)=f(x+\delta)$, $Sf(x)=f(x+\gamma)$. 
For simplicity of display, we also write 
$\widehat{f}=Tf(x)=f(x+\delta),
\overline{f}=Sf(x)=f(x+\gamma),\underline{f}=S^{-1}f(x)=f(x-\gamma)$ and so on.
Then we can write (\ref{Tff}) as
\begin{eqnarray}
\mathsf{ T}(f-\widehat{f})=(1-S)(1-TS^{-1})f,
\end{eqnarray}
for example. Setting $g=(1-T)f$, this `formally' can be expressed as
\begin{eqnarray}
\mathsf{ T}g={(1-S)(1-TS^{-1})\over 1-T}g.
\end{eqnarray}

\begin{prop}
Suppose that $w(z)$ is holomorphic on $D$ and satisfies the periodicity
$w(x+1)=w(x)$. Set
$\eta(x)=w(x)-w(x+\delta)+\eta_0$, where 
$\eta_0= \int_{-1/2}^{1/2}\eta(x)dx$ denotes the zero Fourier component.
Then we can recast 
(\ref{ILW}) into the difference equation
\bb
\frac{\partial}{\partial t}(w-\widehat{w}+\eta_0)
=(w-\widehat{w}+\eta_0)
(w-\overline{w}
-\widehat{\underline{w}}+\widehat{w}).
\label{eqDD1}
\ee
\end{prop}
\begin{rem}
Note that we have $ {d\eta_0/ dt}=0$ from the assumption on $w(x)$.
\end{rem}

\begin{prop}
Let
$\varepsilon$ and $\eta_0$ be constants.
Assume that $\tau$
satisfies the bilinear equation 
\bb
D_t\widehat{\tau}\cdot\tau=
\varepsilon\underline{\widehat{\tau}}\overline{\tau}-\eta_0\widehat{\tau}\tau,
\label{bilinear}
\ee
where $D_t$ denotes the Hirota derivative defined by $D_t f\cdot g=(\partial_t f)g-f(\partial_t g)$.
Set
$\ds w=-\frac{\partial_t\tau}{\tau}$. Then 
$w$ satisfies 
the difference equation (\ref{eqDD1}).
\end{prop}
\proof From (\ref{bilinear}) we have
\begin{eqnarray*}
w-\widehat{w}+\eta_0=
-{\partial_t \tau\over \tau}+
{\partial_t \widehat{\tau}\over \widehat{\tau}}+\eta_0=
\varepsilon {\widehat{\underline{\tau}}\,\overline{\tau}\over \widehat{\tau}\, \tau},
\end{eqnarray*}
hence 
\begin{eqnarray*}
{\partial_t} \log(w-\widehat{w}+\eta_0)=
-{\partial_t \tau\over \tau}
+{\partial_t \overline{\tau}\over \overline{\tau}}
+{\partial_t \widehat{\underline{\tau}}\over \widehat{\underline{\tau}}}
-{\partial_t \widehat{\tau}\over \widehat{\tau}}
=
(w-\overline{w}
-\widehat{\underline{w}}+\widehat{w}).
\end{eqnarray*}
\qed

\subsection{some special solutions}
We give some examples of special solutions to 
(\ref{ILW}), (\ref{eqDD1}), or (\ref{bilinear}).

\begin{prop}\label{tau-1}
Let 
$n$ be a positive integer. Let  $c_1,\cdots,c_n$ be $n$ complex parameters,
$k_1,\cdots k_n$ be $n$ integers which are all distinct and nonzero. 
Set 
\begin{eqnarray}
&&\tau=\det (f_{l,m})_{1\leq l,m\leq n},\label{tau-soliton}\\[1mm]
&&f_{l,m}=\lambda_l^{m-1}+c_l \mu_l^{m-1}
\exp\bigl(2\pi i k_l x
+
(\mu_l-\lambda_l)t
  \bigr),\nonumber\\
&&\lambda_l=-\varepsilon{1-e^{2 \pi i (\delta-\gamma)k_l}\over 
1-e^{2 \pi i \delta k_l}},
\qquad
\mu_l=-\varepsilon{1-e^{-2 \pi i (\delta-\gamma)k_l}\over 
1-e^{-2 \pi i \delta k_l}}.\nonumber
\end{eqnarray}
Then this $tau$ satisfies the bilinear equation
(\ref{bilinear}) written for $\eta_0=\varepsilon$.
\end{prop}

The proof of Proposition \ref{tau-1} will be given in Section \ref{Casorati}.

\begin{rem}
We have
\begin{eqnarray}
\mu_l-\lambda_l=\varepsilon{(1-e^{2 \pi i \gamma k_l})(1-e^{2 \pi i (\delta-\gamma)k_l})
\over 
1-e^{2 \pi i \delta k_l}}.
\end{eqnarray}
\end{rem}

Now we turn to the case of an elliptic solution.
Let 
$\Delta$ be a complex number satisfying ${\rm Im}(\Delta)>0$ and define
\begin{eqnarray}
&&\vartheta_1(u,\Delta)=
-i e^{\pi  i \Delta/4+\pi i u}
\sum_{m\in {\bf Z}}(-1)^m e^{\pi i \Delta m(m+1)+2\pi i mu}.
\end{eqnarray}

\begin{prop}\label{theta-sol}
Let 
$\varepsilon$ and $\Delta$ be a complex parameters satisfying 
${\rm Im}(\Delta)>{\rm Im}(\delta)$, and let 
$k$ be a nonzero integer. Set
\begin{eqnarray}
&&\tau=\vartheta_1(kx+\omega(k) t,\Delta),\\
&&\omega(k)=-\varepsilon 
{\vartheta_1(k\gamma,\Delta)\vartheta_1(k(\delta-\gamma),\Delta)
\over \vartheta_1'(0,\Delta)\vartheta_1(k\delta,\Delta)},
\end{eqnarray}
and 
\begin{eqnarray}
&&\eta_0=
\omega(k)\left(
{\vartheta_1'(k(\gamma-\delta),\Delta)\over \vartheta_1(k(\gamma-\delta),\Delta)}-
{\vartheta_1'(k\gamma,\Delta)\over \vartheta_1(k\gamma,\Delta)}\right).
\end{eqnarray}
Then $\tau$ satisfies (\ref{bilinear}) and 
\begin{eqnarray}
\eta=\varepsilon {\widehat{\underline{\tau}}\, \overline{\tau}\over \widehat{\tau}\,\tau},
\end{eqnarray}
is a special solution to the integro-differential equation (\ref{ILW}).
\end{prop}

It is straightforward to check this by using addition formulas, so we omit the proof.

\begin{rem}
Note that, the $\Delta$ (not the $\delta$ of the period of the equation (\ref{ILW}))
gives the period of the solution.
\end{rem}

\subsection{limit to periodic ILW}

Now we consider the limit $\gamma \rightarrow 0$, and derive the periodic ILW equatin 
studied in \cite{AFSS}. 
Let  ${\cal T}$ be the integral operator 
\bb
({\cal T}f)(x)
=\
{i\omega_1\over \pi }
\int_{-1/2}^{1/2}
\hspace{-2.2em}\backslash\hspace{.5em}\,\,\,\,\,\left\{
\zeta(2\omega_1(y-x))-2\zeta(\omega_1)(y-x)\right\}\cdot
f(y)dy,
\label{sekibunhenkan}
\ee
then we have 
\begin{eqnarray}
{\cal T}(e^{2\pi i m x})=-{1\over 2}{1+p^m\over 1-p^m}e^{2\pi i m x}\qquad (m\neq 0),
\end{eqnarray}
and ${\cal T}(1)=0$. Hence we have the formal expression 
\begin{eqnarray}
{\cal T}g=-{1\over 2}{1+T\over 1-T}g.
\end{eqnarray}

\begin{lem}
We have the expansion of $\mathsf{T}$ in $\gamma$ as
\begin{eqnarray}
\mathsf{ T}=-\gamma \partial_x +\gamma^2{\cal T}\partial^2+O(\gamma^3).
\end{eqnarray}
\end{lem}

To have the periodic ILW equation, we need to cancel the leading term $-\gamma \partial_x $. 
To this end we first perform a Galilean  transformation on $\eta(x,t)$ and assume
the expansion as $e^{ct\partial_x}\eta=\varepsilon+\gamma u(x,t)+O(\gamma^2)$, where $c$ and $\varepsilon$ are constants.
Rescaling $t$ as 
$t\rightarrow \gamma^{-2} t$ and setting $c=\varepsilon \gamma+a\gamma^2+\cdots$,
we have the integro-differential equation of ILW-type for $u$
\begin{eqnarray}
u_t=a u_x-uu_x+\varepsilon {\cal T} (u_{xx}).
\end{eqnarray}

\section{Lax formalism}
\subsection{Sato theory}
We introduce an infinite set of independent variables
$x$, $r$ and 
${\bf t}=(t_1=t,t_2,t_3,\cdots)$. 
Define difference operators $S,R$ and $T$ by
$Sf(x,r,{\bf t})=f(x+\gamma,r,{\bf t})$,
$Rf(x,r,{\bf t})=f(x,r+\delta,{\bf t})$,
$Tf(x,r,{\bf t})=f(x+\delta,r,{\bf t})$. For simplicity, 
we write $Sf=\overline{f}$, $Rf=\widetilde{f}$, 
$Tf=\widehat{f}$ and so on.

We formulate a version of Sato theory 
based on the papers \cite{UT} and \cite{TS}.
In what follows, we work with a space of operators expressed as formal series in $S^{-1}$.
Let $W=1+w_1S^{-1}+w_2S^{-2}+\cdots$ be the Sato-Wilson operator. 
Let $\phi(\lambda)$ and $\kappa(\lambda)$ be Laurent series in 
$\lambda$ as
\begin{eqnarray}
\phi(\lambda)=\lambda+\sum_{i\geq 0} \phi_{i+1} \lambda^{-i},\qquad
\kappa(\lambda)=\lambda+\sum_{i\geq 0} \kappa_{i+1} \lambda^{-i},\qquad 
\phi_i,\kappa_i\in{\bf C},
\end{eqnarray}
and define $\rho(\lambda)=\phi^{-1}(\lambda)$ by the condition
\begin{eqnarray}
\rho(\lambda)=\lambda+\sum_{i\geq 0} \rho_{i+1} \lambda^{-i},\qquad
\phi(\rho(\lambda))=\lambda.
\end{eqnarray}

We impose the following set of evolution equations on $W$:
\bb
&&\ds\frac{\partial W}{\partial t_j}
+W\phi(S)^j=B_jW,\qquad B_j=(W\phi(S)^jW^{-1})_+,\label{hatten-1}\\
&&\ds\widetilde{W}\kappa(S)=CW,\qquad C=(\widetilde{W}\kappa(S) W^{-1})_+,
\label{hatten}
\ee
where we have used the standard notation 
that we write $(A)_+=\sum_{j=k}^0 a_jS^{-j}$
for any $A=\sum_{j=k}^\infty a_jS^{-j}$.

\begin{rem}
The equality 
(\ref{hatten}) means the truncation of the operator
$\widetilde{W}\kappa(S)W^{-1}=(\widetilde{W}\kappa(S) W^{-1})_+$,
which involves an infinite system of equations about integrals of motion,
and is very much important for our task.
Namely, the $\kappa_m$'s plays the role of integrals.
On the other hand, the parameters $\phi_m$'s are 
introduced just to have some linear combination of 
the time variables $t_i$'s.
See Section \ref{int-1} and Section \ref{int-2}.
\end{rem}

\noindent
{\bf Example of the time evolution 
(\ref{hatten-1})}: We have
\begin{eqnarray}
&&\partial_{t_1}w_k=w_k(w_1-\overline{w}_1)-(w_{k+1}-\overline{w}_{k+1})
-\phi_{k+1}-\sum_{j=1}^{k-1}\phi_{j+1}w_{k-j},
\end{eqnarray}
etc.

\begin{prop}
The compatibility among the evolutions in the ${\bf t}$ directions and the $r$ directions
can be described by the 
Zakharov-Shabat equations
\bb
&&{\partial B_i\over \partial t_j}-{\partial B_j\over \partial t_i}+
[B_i,B_j]=0,\\
&&\frac{\partial C}{\partial t_i}
+CB_i-\widetilde{B}_iC=0.
\label{eqZS}
\ee
\end{prop}

Introduce the wave function $\Psi$ defined by
\begin{eqnarray}
&&\Psi(x,r,{\bf t})= W \rho(\lambda)^{x/\gamma}\kappa(\lambda)^{r/\delta} e^{\xi({\bf t},\lambda)},\\
&&\xi({\bf t},\lambda)=\sum_{i=1}^\infty t_i \lambda^i.
\end{eqnarray}
\begin{prop}
The wave function $\Psi$ satisfies 
\begin{eqnarray}
&&{\partial\Psi\over \partial t_i}=B_i \Psi,\\
&&\widetilde{\Psi}=C\Psi.
\end{eqnarray}
\end{prop}

The operators 
$B_1$ and $C$ can be written explicitly as
\bb
B_1=S+w_1-\overline{w}_1+\phi_1,\quad 
C=S+\widetilde{w}_1-\overline{w}_1+\kappa_1.
\label{eqw1}
\ee
Write $w=w_1$ for simplicity, then from Zakharov-Shabat equation (\ref{eqZS}) we have
\bb
{\partial\over \partial t}(\widetilde{w}-\overline{w}+\kappa_1)
+
(w-\widetilde{w}-\overline{w}+\widetilde{\overline{w}})
(\widetilde{w}-\overline{w}+\kappa_1)=0.
\ee

\begin{dfn}
We call the condition on $W$
\bb
W(x,r+\delta,{\bf t})=W(x+\gamma-\delta,r,{\bf t}),
\label{reductioncondition}
\ee
($\widetilde{W}=T^{-1}S W T S^{-1}$
or $\widehat{\widetilde{\underline{w}}}_k=w_k$)  
the reduction condition.
\end{dfn}

\begin{prop}
Assume that 
$W$ satisfies the reduction condition (\ref{reductioncondition}), and 
define the Lax operator by $L=TS^{-1}C=
WTS^{-1}\kappa(S)W^{-1}$. Setting
$\eta_0=\kappa_1$, $\eta=w-\widehat{w}+\eta_0$, 
we have $L=T+\eta TS^{-1}$, and 
$w$ satisfies the difference equation
(\ref{eqDD1}). Hence 
$\eta$ satisfies (\ref{ILW}) under the condition that 
$w$ is holomorphic on $D$ and periodic $w(x+1)+w(x)$.
\end{prop}

\begin{prop}
Assuming the 
reduction condition (\ref{reductioncondition}), we have
\begin{eqnarray}
&&{\partial\Psi\over \partial t_i}=B_i \Psi,\\
&&L\Psi=\rho(\lambda)^{(\delta-\gamma)/\gamma}\kappa(\lambda)\Psi.\label{Lax-}
\end{eqnarray}
\end{prop}

\subsection{Casorati determinant}\label{Casorati}
As an application of the Sato theory we have studied in the previous subsection, 
we construct a special solution of (\ref{ILW}) in terms of a Casorati determinant.
In this subsection we restrict ourself to the simplset possible situation 
$\phi(\lambda)=\rho(\lambda)=\lambda$ and $\kappa(\lambda)=\lambda+\varepsilon$
and assume the truncation of the Sato-Wilson operator as
\begin{eqnarray}
&&W=1+w_1S^{-1}+w_2S^{-2}+\cdots+w_nS^{-n}.
\end{eqnarray}
We note that the case  $\varepsilon=0$ necessarily gives us the  
trivial situation $\eta=0$.
For simplicity we sometimes denote $S^k f(x)=f^{(k)}(x)$ etc.

Consider the linear system $WS^nf=0$. Let 
$f_1,f_2,\cdots,f_n$ be the basis of the linear system satisfying
the dispersion relation
\begin{eqnarray}
&&\ds\frac{\partial f_i}{\partial t_j}
=S^jf_i,\\
&&\ds Rf_i=(S+\varepsilon)f_i.
\label{eqDis}
\end{eqnarray}
As usual, we explicitly give the basis as 
\begin{eqnarray}
&&f_j=\exp\left({x\over \gamma} 
\log \lambda_j+{r\over \delta}\log(\lambda_j+\varepsilon) +
\sum_{k\geq 1}\lambda_j^k  t_k\right)\\
&&\qquad +c_j
 \exp\left({x\over \gamma} 
\log \mu_j+{r\over \delta}\log(\mu_j+\varepsilon) +
\sum_{k\geq 1}\mu_j^k  t_k\right),\nonumber
\end{eqnarray}
by introducing the set of parameters 
$\{\lambda_i,\mu_i,c_i|1\leq i\leq n\}$.
\begin{prop}
The operator $W$ uniquely characterized by the linear system 
$W S^n f_i$  ($i=1,\cdots,n$)
satisfies the evolution equations (\ref{hatten-1}),(\ref{hatten}) written for 
$\phi(\lambda)=\lambda,\rho(\lambda)=\lambda,
\kappa(\lambda)=\lambda+\varepsilon$.
\end{prop}

For any sequence of integers
$\alpha_1,\alpha_2,\cdots,\alpha_n$, we denote the correspondong 
Casorati determinant by the symbol
$|\alpha_1,\alpha_2,\cdots,\alpha_n|=
\det(f_i^{(\alpha_j)})_{1\leq i,j\leq n}$. 
Setting $\tau=|n-1,n-2,\cdots,0|$, 
the $w_i$'s are written as
\bb
w_k=(-1)^k |n,n-1,\cdots,\check{k},\cdots,0|/\tau,\label{eqCra}
\ee
where the symbol $\check{k}$ means that the letter $k$ is eliminated.
Introduce differential operators $p_k$ by
\begin{eqnarray}
\sum_{k=0}^\infty p_k y^k=
\exp\left(-\sum_{m=1}^\infty {1\over m}\partial_{t_m} y^m\right).
\end{eqnarray}

\begin{prop}
We have $w_k=(p_k \tau)/\tau$, namely 
\begin{eqnarray}
\sum_{i=0}^\infty  w_k(x,r,{\bf t}) \lambda^{-i}=
{\tau(x,r,{\bf t}-[\lambda^{-1}])\over 
\tau(x,r,{\bf t})},
\end{eqnarray}
where we have used the standard notation 
$[\lambda]=(\lambda/1,\lambda^2/2,\lambda^3/3,\cdots)$. 
\end{prop}

Note that we especially have 
$w_1=-(\partial_{t_1}\tau)/\tau$, 
$w_n=(-1)^n(S\tau)/\tau$ and $w_{n+1}=w_{n+2}=\cdots=0$. Hence we have 
\bb
B_1=S+\frac{\partial_t w_n}{w_n},\quad 
C=S+\varepsilon\frac{\widetilde{w}_n}{w_n}.
\label{eqwn}
\ee

\begin{prop}
If the condition 
\begin{eqnarray}
{\lambda_j+\varepsilon\over\mu_j+\varepsilon}=
(\lambda_j/\mu_j)^{(\gamma-\delta)/\gamma},
\end{eqnarray}
is satisfied, the 
reduction condition (\ref{reductioncondition}) holds. 
Such a pair $\lambda_j,\mu_j$ is parametrized by 
via $k_j$ as
\begin{eqnarray}
\lambda_j=-\varepsilon{1-e^{2 \pi i (\delta-\gamma)k_j}\over 
1-e^{2 \pi i \delta k_j}},
\quad 
\mu_j=-\varepsilon{1-e^{-2 \pi i (\delta-\gamma)k_j}\over 
1-e^{-2 \pi i \delta k_j}}.
\end{eqnarray}
\end{prop}

Assuming the reduction condition (\ref{reductioncondition}) and setting
$L=W(T+\varepsilon TS^{-1})W^{-1}=T+\eta TS^{-1}$, 
we have the representation of the $\eta$ which satisfies (\ref{ILW}) in terms of 
the $\tau$ in two ways as
\begin{eqnarray}
&&\eta=-{\partial_t \tau\over \tau}+
{\partial_t \widehat{\tau}\over \widehat{\tau}}+
\varepsilon=\varepsilon
{\widehat{\underline{\tau}}\,\,\overline{\tau}
\over \widehat{\tau}\,\,\tau}.
\end{eqnarray}
This means that the $\tau$ satisfies the bilinear equation (\ref{bilinear}). 
Thus we have the special solution presented in Proposition \ref{tau-1}.

The wave function $\Psi$ associated with this solution reads
\begin{eqnarray}
\Psi={\tau(x,r,{\bf t}-[\lambda^{-1}])\over 
\tau(x,r,{\bf t})}\lambda^{x/\gamma}
(\lambda+\varepsilon)^{r/\delta}e^{\xi({\bf t},\lambda)},
\end{eqnarray}
which satisfies (under the condition (\ref{reductioncondition}))
\begin{eqnarray}
L\Psi=\lambda^{(\delta-\gamma)/\gamma}(\lambda+\varepsilon)\Psi.
\end{eqnarray}

\subsection{conservation laws}\label{int-1}
Throughout this subsection, we assume that the reduction condition (\ref{reductioncondition}) is satisfied.
Introduce the Fourier expansions
\begin{eqnarray}
&&\eta(x)=\sum_{n\in {\bf Z}} \eta_{-n}e^{2\pi i nx},\qquad
w_k(x)=\sum_{n\in {\bf Z}} w_{k,-n}e^{2\pi i nx},\\
&&\eta_n=\int_{-1/2}^{1/2}dx \,\eta(x) e^{2\pi i n x},\qquad
w_{k,n}=\int_{-1/2}^{1/2}dx \,w_k(x) e^{2\pi i n x}.
\end{eqnarray}

\begin{con}\label{kappa-eta}
Assume the evolution equation in the $r$ direction 
(\ref{hatten}) and the 
reduction condition (\ref{reductioncondition}).
The 
$\kappa_k$'s are given as certain degree $k$ expressions in 
$\{\eta_n\}$. In other words, 
$\eta(x)$ satisfies infinitely many constraint conditions 
for given $\{\kappa_k\}$.
\end{con}

\noindent
{\bf Example:}
\begin{eqnarray}
&&\kappa_1=\eta_0,\label{kappa_1}\\
&&\kappa_2=\sum_{n\neq 0}{p^nq^{-n}\over 1-p^n}\eta_n\eta_{-n},\\
&&\kappa_3=\sum_{m\neq 0,n\neq 0}
{p^mq^{-m}\over 1-p^m}{p^nq^{-n}\over 1-p^n}
\eta_{m}\eta_{-m+n}\eta_{-n}-
\sum_{n\neq 0}{p^nq^{-n}\over (1-p^n)^2}\eta_0\eta_n\eta_{-n},\label{kappa_3}
\end{eqnarray}
etc.

We present some explicit calculations to explain what is meant by Conjecture \ref{kappa-eta}.
\begin{lem} It follows from $L=T+\eta TS^{-1}=WTS^{-1}\kappa(S)W^{-1}$ that
\begin{eqnarray}
\eta \widehat{\underline{w}}_{k-1}=w_k-\widehat{w}_k
+\sum_{j=0}^{k-1}\kappa_{j+1} w_{k-j-1}. \label{w-eta-kappa}
\end{eqnarray}
\end{lem}

First we consider the case  $k=1$ in  (\ref{w-eta-kappa}), namely  $\eta=w_1-\widehat{w}_1+\kappa_1$. From this we have
the relations among the Fourier modes of $w_1,\eta$ and $\kappa_1$ as
\begin{eqnarray}
&&w_{1,-n}={1\over 1-p^n}\eta_{-n}\qquad (n\neq 0),\\
&&\kappa_1=\eta_0,\\
&&w_{1,0}\mbox{ is free}.
\end{eqnarray}
Nextly, by setting $k=2$ (\ref{w-eta-kappa}) gives us
$\eta\widehat{\underline{w}}_1=w_2-\widehat{w}_2+\kappa_1w_1+\kappa_2$. Therefore we have
\begin{eqnarray}
&&w_{2,-n}={1\over 1-p^n}
\left( 
\sum_{l\neq 0} {p^lq^{-l}\over 1-p^l}
\eta_{-n+l}\eta_{-l}+\eta_{-n}w_{1,0}-
\kappa_1 {1\over 1-p^n}\eta_{-n}\right)
\qquad (n\neq 0),\\
&&\kappa_2=
\sum_{l\neq 0} {p^lq^{-l}\over 
1-p^l}
\eta_{l}\eta_{-l},\\
&&w_{2,0}\mbox{ is free},
\end{eqnarray}
and so on. In this way, we obtain a series of constraints (\ref{kappa_1})-(\ref{kappa_3}) etc, besides
the relations giving the nonzero Fourier modes of $w_k$'s in terms of the $\eta_k$'s.

\begin{con}
For $k=1,2,\cdots$ and nonzero integer $n$, 
$w_{k,-n}$ is expressed in terms of the $\eta_k$'s and $w_{1,0},\cdots,w_{k-1,0}$.
\end{con}

Now we present a conjecture that the constraints (\ref{kappa_1})-(\ref{kappa_3}) etc 
can be written as sums of  multiple integrals whose kernel is simply given by
products of theta functions with the period $\delta$:
\begin{eqnarray}
\vartheta_1(x;\delta)&=&-i p^{1/8}e^{\pi i x}
\sum_{n\in {\bf Z}}(-1)^n p^{n(n+1)/2}e^{2 \pi i nx}\\
&=&i p^{1/8}e^{-\pi i x}\prod_{n\geq 1}
(1-p^{n-1} e^{2 \pi i x})(1-p^{n}e^{-2 \pi i x})(1-p^{n}).\nonumber
\end{eqnarray}
Note that we have $\vartheta_1'=\vartheta_1'(0)=2 \pi p^{1/8}\prod_{n\geq 1}(1-p^{n})^3$ and
$\vartheta_1(-x)=-\vartheta(x)$.

\begin{dfn}\label{def-I_n}We define the quantities $I_n$ ($n=1,2,\cdots$) by the multiple integral
\begin{eqnarray}
I_n&=&(-1)^{n-1}{1\over n!}{(\vartheta_1' )^{n-1}\vartheta_1(n\gamma)\over 
(2\pi i)^{n-1}
\vartheta_1(\gamma)^n}
\int_{-1/2}^{1/2}\cdots \int_{-1/2}^{1/2}dx_1\cdots dx_n\times \label{I_n}\\
&&\times \prod_{1\leq i<j\leq n}
{\vartheta_1(x_i-x_j)^2\over 
\vartheta_1(x_i-x_j+\gamma)\vartheta_1(x_i-x_j-\gamma)}\cdot 
\eta(x_1)\cdots \eta(x_n),\nonumber
\end{eqnarray}
where we have used the theta function with the period $\delta$ and denoted $\vartheta_1(x)=\vartheta_1(x;\delta)$
for short.
\end{dfn}

\begin{rem}
An explanation is in order here. Similar multiple integrals as in (\ref{I_n})
were found in \cite{S} when one of the authors studied the 
family of Macdonald difference operators in terms of the free field construction. 
In \cite{FHMSWY}, it is found that the algebra found by Feigin and Odesskii \cite{FO}
is the underlying algebraic structure for such 
integrals, including some elliptic extension of the Macdonald theory.  
{}From this point of view, one can regard the integrals (\ref{I_n})
as a certain classical limit of this. Some detail will be given in Section \ref{Poisson}.
\end{rem}

\begin{con}
Let the degree of $\kappa_k$ be $k$.
The $I_n$'s are homogeneous polynomials of degree $n$ in the $\kappa_k$'s.
Namely the quantities $I_n$'s are conserved under the time evolution described by 
the integro-differential equation (\ref{ILW}).
\end{con}

\noindent
{\bf Examples:}
\begin{eqnarray}
&&I_1=\kappa_1,\\
&&I_2=\kappa_2+
{1\over 2}\left(\sum_{n\neq 0}{q^n-p^nq^{-n}\over 1-p^n}\right)\kappa_1^2,\\
&&I_3=\kappa_3+
\left(\sum_{n\neq 0}{q^n-p^nq^{-2n}\over 1-p^n}\right)\kappa_1\kappa_2+\\
&&\qquad +{1\over 6}\left(
\sum_{m,n\neq 0}{q^m-p^mq^{-2m}\over 1-p^m}{q^n-p^nq^{-n}\over 1-p^n}
-2
\sum_{n\neq 0}{q^n(q^n-p^nq^{-2n})\over (1-p^n)^2}\right)\kappa_1^3,\nonumber
\end{eqnarray}
and so on.

\subsection{Conserved densities}\label{int-2}
In this subsection, we continue our study on structure of the 
conserved quantities from the point of view of conserved densitiess.

First let us study the operator $WS^kW^{-1}$ in some detail. 
Set
\begin{eqnarray}
&&WS^kW^{-1}=S^k+\sum_{l=1}^\infty u_{k,l}S^{k-l}.
\end{eqnarray}
\begin{lem} Write 
$\omega_0=1,u_{k,0}=1$ for simplicity. Then we have
\begin{eqnarray}
\sum_{l=0}^p u_{k,l}w_{p-l}^{(k-l)}=w_p.
\end{eqnarray}
Hence $u_{k,l}$ can be expressed as the determinant
\begin{eqnarray}
u_{k,l}=
\left|
\begin{array}{ccccc}
1&0 & \cdots&0 &1\\
w_1^{(k)}&1&&&w_1\\
w_2^{(k)}&w_1^{(k-1)}&\ddots&&\vdots\\
\vdots &\vdots&&1&w_{l-1}\\
w_l^{(k)}&w_{l-1}^{(k-1)}&\cdots&w_1^{(1)}&w_l\\
\end{array}
\right|. \label{u_kl}
\end{eqnarray}
\end{lem}

As usual, we write ${\rm res}\,A=a_0$ for any
$A=\sum_{j=k}^\infty a_jS^{-j}$. 
\begin{prop}
For any positive integer $k$, $u_{k,k}={\rm res}\,(WS^kW^{-1})$ is a total $q$-difference. 
\end{prop}
\proof 
For $j=1,2,3\cdots$, set
\begin{eqnarray*}
\xi_j=
\left|
\begin{array}{ccccc}
w_1^{(j)}&1&0&\cdots & 0\\
w_2^{(j)}&w_1^{(j-1)}&\ddots&&\vdots\\
\vdots&\vdots&\ddots&\ddots&0\\
\vdots &\vdots&&\ddots&1\\
w_j^{(j)}&w_{j-1}^{(j-1)}&\cdots&&w_1^{(1)}\\
\end{array}
\right|,
\end{eqnarray*}
and set $\xi_0=1$. With this notation we can write the determinant (\ref{u_kl}) written for $l=k$ as
\begin{eqnarray*}
u_{k,k}=(-1)^k\sum_{j=1}^k (-1)^j \left(w_j\xi_j-w_j^{(j)}\xi_j^{(j)} \right).
\end{eqnarray*}
\qed

Here are some examples of $u_{k,l}$:
\begin{eqnarray}
&&u_{1,1}={\rm res}\,(WSW^{-1})=(1-S)w_1,\\
&&u_{2,1}=(1-S^2)w_1,\\
&&u_{2,2}=
{\rm res}\, (WS^2W^{-1})=
(1-S)(w_2+\overline{w}_2-w_1\overline{w}_1),\\
&&u_{3,1}=(1-S^3)w_1,\\
&&u_{3,2}=
w_2-\overline{\overline{\overline{w}}}_2+
\overline{\overline{w}}_1\overline{\overline{\overline{w}}}_1-
w_1\overline{\overline{w}}_1,\\
&&u_{3,3}=
{\rm res}\,(WS^3W^{-1})=\\
&&\qquad =
(1-S)(w_3+\overline{w}_3+\overline{\overline{w}}_3
-\overline{w}_1w_2-\overline{\overline{w}}_1\overline{w}_2-w_1\overline{\overline{w}}_2
+w_1\overline{w}_1\overline{\overline{w}}_1),\nonumber
\end{eqnarray}
etc.

Now we can state the structure of the residue of 
$B_k=(W\phi(S)^kW^{-1})_+$.
\begin{prop}\label{E_k}
For any positive integer $n$, there exists a difference polynomial $E_k$ of the $w_k$'s
such that ${\rm res}\, B_n=const.+(1-S)E_n$.
\end{prop}

\noindent
{\bf Examples:}
\begin{eqnarray}
&&{\rm res}\,B_1=\phi_1+(1-S)w_1,\\
&&
{\rm res}\, B_2=(\phi_1^2+2 \phi_2)+
2\phi_1 (1-S)w_1+
(1-S)(w_2+\overline{w}_2-w_1\overline{w}_1),\\
&&
{\rm res}\, B_3=
(\phi_1^3+6\phi_1\phi_2+3 \phi_3)+
3(\phi_1^2+\phi_2)(1-S)w_1+
3 \phi_1(1-S)(w_2+\overline{w}_2-w_1\overline{w}_1)+\nonumber\\
&&\qquad+
(1-S)(w_3+\overline{w}_3+\overline{\overline{w}}_3
-\overline{w}_1w_2-\overline{\overline{w}}_1\overline{w}_2-w_1\overline{\overline{w}}_2
+w_1\overline{w}_1\overline{\overline{w}}_1),
\end{eqnarray}
and so on.

Now we assume the reduction condition (\ref{reductioncondition}).

\begin{lem}From the evolution equation $\partial_{t_m}L=[B_m,L]$, we have
\begin{eqnarray}
\partial_{t_m}\eta=\eta \cdot (1-TS^{-1}){\rm res}\, B_m.
\end{eqnarray}
Hence from Proposition \ref{E_k}, we have
\begin{eqnarray}
\partial_{t_m}\eta=\eta \cdot (1-TS^{-1})(1-S)E_m.\label{kore}
\end{eqnarray}
\end{lem}

Suppose that the $E_m$'s are holomorphic on $D$ and 
periodic $E_m(x+1)=E_m$. Then from Corollary \ref{cor} and (\ref{kore}) we have some quantity $H_m$
which satisfies $\partial_{t_m}\eta=\eta \cdot \mathsf{T}(H_m)$.
We have a conjecture which explicitly gives $H_m$.
\begin{dfn}\label{def-H_n}We define the densities $H_n(x)$ ($n=1,2,\cdots$) by the $n-1$-fold multiple integral
\begin{eqnarray}
H_n(x_1)&=&(-1)^{n-1}{1\over (n-1)!}{(\vartheta_1' )^{n-1}\vartheta_1(n\gamma)\over 
(2\pi i)^{n-1}
\vartheta_1(\gamma)^n}
\int_{-1/2}^{1/2}\cdots \int_{-1/2}^{1/2}dx_2\cdots dx_n\times \label{H_n}\\
&&\times \prod_{1\leq i<j\leq n}
{\vartheta_1(x_i-x_j)^2\over 
\vartheta_1(x_i-x_j+\gamma)\vartheta_1(x_i-x_j-\gamma)}\cdot 
\eta(x_1)\cdots \eta(x_n),\nonumber
\end{eqnarray}
namely we have $n I_n=\int_{-1/2}^{1/2}dx H_n(x)$.
\end{dfn}
\begin{con}\label{m-th-order}
By suitably choosing the constants $\phi_k$'s and $c$, we have
the equality $H_m=E_m-\widehat{E}_m+c$, namely we have
\begin{eqnarray}
\partial_{t_m}\eta=\eta \cdot \mathsf{T}(H_m).
\end{eqnarray}
\end{con}

We show some examples.
Firstly, we have
$E_1=\omega_1$, and 
$H_1=w_1-\widehat{w}_1+c$. By choosing the constant as 
$c=\kappa_1$ we have $H_1=\eta$.
Nextly, we have $E_2=w_2+\overline{w}_2-w_1\overline{w}_1+2\phi_1 w_1$. Thus 
choosing the constants as 
\begin{eqnarray}
&&c=2\kappa_2-\kappa_1(\kappa_1-2\phi_1),\\
&&2\phi_1=\kappa_1\left(1+\sum_{n\neq 0}{q^n-p^nq^{-n}\over 1-p^n}\right),
\end{eqnarray}
we have
\begin{eqnarray}
E_2-\widehat{E}_2+c
=-(w_1-\widehat{w}_1+\kappa_1)
(\overline{w}_1-\widehat{\underline{w}}_1+\kappa_1-2\phi_1)=H_2.
\end{eqnarray}

\section{Poisson structure}\label{Poisson}
In this section, we study a Poisson structure 
which is obtained as a certain deformation
of the algebra associated with the family of Macdonald difference operators $D_n^r(q,t)$ 
(see $(3.4)_r$ in VI.3 of \cite{Mac}),
and study the relations with the integro-differential equation (\ref{ILW})
and the Lax formulation which we have developed in the previous section.

One of the authors studied \cite{S}
the family of Macdonald operators 
in terms of  a Heisenberg algebra and its Fock representation (namely in the case of
infinitely many variables $x_1,x_2,\cdots$). In \cite{FHMSWY} is given 
a systematic description of the relation between this and the algebra obtained by Feigin and Odesskii \cite{FO}
(whose trigonometric limit to be very precise). 
In other words, we analyzed the commuting family of operators introduced by Macdonald from the point of view of
a pairing between the two kinds of quantum i.e. noncommutative algebras.

Once we note that the algebra of Feigin and Odesskii is originally
constructed over an elliptic curve (which contains three parameters in this
most general setting) one may easily find the corresponding elliptic  deformation of 
the commutative family which are acting on the Fock space containing three parameters, say $q$, $t$ and $p$. 

It is interesting to note that these underlying algebras, the Heisenberg algebra and the Feigin-Odesskii algebra,
become commutative in the limit $t\rightarrow 1$ with $q$ and $p$ fixed. Hence by setting $t=e^\hbar$ and considering the limit $\hbar \rightarrow 0$,
one may naturally define Poisson algebras on the corresponding commutative algebras.
\bigskip

Because we lack the space, we skip the derivation and just give the
resulting Poisson algebra, then compare the relations with those coming from the Lax formalizm.
We again use the notations $q=e^{2\pi i \gamma},p=e^{2\pi i \delta}$  for simplicity of display.
The Poisson algebra we study is generated by 
$\{\lambda_n|n\in{\bf Z}\setminus \{0\}\}$
with the Poisson bracket
\begin{eqnarray}
\{\lambda_n,\lambda_m\}=
{(1-q^n)(1-p^nq^{-n})\over 1-p^n}\delta_{n+m,0},
\end{eqnarray}
where $\delta_{m,n}$ denotes the Kronecker delta.
Let 
$\varepsilon$ be a constant and set
\begin{eqnarray}
\eta(x)=\sum_{n\in{\bf Z}}\eta_n e^{-2\pi i nx}=
\varepsilon
\exp\left(\sum_{n\neq 0}\lambda_n e^{-2\pi i nx }\right).\label{eta}
\end{eqnarray}

\begin{prop}
We have 
\begin{eqnarray}
\{\eta(x),\eta(y)\}=
\sum_{n\neq 0}{(1-q^n)(1-p^nq^{-n})\over 1-p^n}
e^{-2\pi i n(x-y)}~\eta(x)\eta(y).
\end{eqnarray}
\end{prop}

Let us define $I_n$ and $H_n(x)$ by the same equation (\ref{I_n}) in Definition \ref{def-I_n} and (\ref{H_n}) in Definition \ref{def-H_n}
respectively 
from the quantity defined by (\ref{eta}). 
Here is a crucial remark: we are not working with the dependent variables 
described by the Lax formalism, but we are starting from 
the Poisson algebra and trying to reconstruct the  same 
hierarchy described by the Lax operator $L$ together with the reduction condition (\ref{reductioncondition}).
First, one can prove the following.
\begin{prop}
For any positive integers $k$ and $l$, we have
\begin{eqnarray}
\{I_k,I_l\}=0.
\end{eqnarray}
\end{prop}

Let $I_n$ be our $n$-th Hamiltonian, and set $\partial \eta/\partial t_n=\{I_n,\eta(x)\}$.
\begin{prop}
We have 
$\partial \eta/\partial t_n=\{I_n,\eta(x)\}=\eta(x) \mathsf{T}(H_n)$.
\end{prop}

For example,  we have
\begin{eqnarray}
{\partial\over \partial t_1}\eta(x)=
\{\eta_0,\eta(x)\}=\eta(x)
\sum_{n\neq 0}{(1-q^n)(1-p^nq^{-n})\over 1-p^n}\eta_{-n}e^{2\pi i nx}.
\end{eqnarray}
The RHS is nothing but $\eta(\mathsf{T}\eta)$, hence 
we recover the integro-differential equation (\ref{ILW}).
Note that the time evolution in general takes the 
same form as we conjecture for the Lax formulation (see Conjecture \ref{m-th-order}).

We note that the $\tau$ also can be presented as a kind of vertex operator.
Set 
\begin{eqnarray}
\tau(x)=
\exp\left(-\sum_{n\neq 0}{p^n\over (1-q^n)(1-p^nq^{-n})}\lambda_n 
e^{-2\pi i nx}\right).
\end{eqnarray}
Then from (\ref{eta}) we have
\begin{eqnarray}
\eta(x)=\varepsilon 
{\widehat{\underline{\tau}}\,\overline{\tau}\over \widehat{\tau}\,\tau}.
\end{eqnarray}
We have
\begin{eqnarray}
\{\eta(x),\tau(y)\}=-\sum_{n\neq 0}{1\over 1-p^n}
e^{-2\pi i n (x-y)} ~\eta(x)\tau(y).
\end{eqnarray}
{} From this we have
\begin{eqnarray}
&&\{\eta(x),\tau(y+\delta)\}/\tau(y+\delta)-
\{\eta(x),\tau(y)\}/\tau(y)
=\left(\delta(y-x)-1\right)\eta(x),
\end{eqnarray}
where we have used the notation $\delta(x)=\sum_n e^{2\pi i n x}$.
This gives us
\begin{eqnarray}
D_1\widehat{\tau}\cdot \tau
=\varepsilon 
\widehat{\underline{\tau}}\overline{\tau}-
 \eta_0 \widehat{\tau} \tau,
\end{eqnarray}
which is exactly the bilinear equation (\ref{bilinear}).

\appendix
\section{Weierstrass $\zeta$ function}

The Weierstrass $\zeta$ function $\zeta(z)=\zeta(z;2\omega_1,2\omega_2)$ is defined by
\begin{eqnarray}
\zeta(z)={1\over z}+{\sum_{m,n}}'
\left\{{1\over z-2m\omega_1-2n\omega_2}+
{1\over 2m\omega_1+2n\omega_2}+
{z\over (2m\omega_1+2n\omega_2)^2}\right\}, \label{zeta}
\end{eqnarray}
where the symbol ${\sum_{m,n}}'$ means
the summation over $(m,n)\in{\bf Z}^2\setminus\{(0,0)\}$.
We have 
\begin{eqnarray*}
&&\zeta(-z)=-\zeta(z),\\
&&\zeta(z+2\omega_1)=\zeta(z)+2\zeta(\omega_1),\\
&&\zeta(z+2\omega_2)=\zeta(z)+2\zeta(\omega_2),
\end{eqnarray*}
and the Fourier expansion
\begin{eqnarray}
\zeta(u)={\zeta(\omega_1)\over \omega_1}u+
{\pi \over 2\omega_1}\cot {\pi u\over 2\omega_1}+
{2\pi \over \omega_1}\sum_{n=1}^\infty {p^{n}\over 1-p^{n}}\sin {\pi n u\over \omega_1},\label{fou}
\end{eqnarray}
where $p=e^{2 \pi i \omega_2/\omega_1}$.

\end{document}